\documentstyle[11pt,aaspp4]{article}

\begin{document}

\title{W~UMa-type Binary Stars in Globular Clusters}

\author{Slavek M. Rucinski}
\affil{David Dunlap Observatory, University of Toronto\\ 
P.O.Box 360, Richmond Hill, Ontario, Canada L4C 4Y6\\
{\it e-mail: rucinski@astro.utoronto.ca}}

\centerline{\today}

\begin{abstract}
A sample of 86 contact binary systems in 14 globular clusters
with available color index data in $(B-V)$ or in $(V-I)$
has been analyzed. A large fraction of all systems (at least
one third) are numerous foreground Galactic Disk projections 
over long lines of sight to the clusters. Since the selection
of the cluster members has been based on the $M_V(\log P, color)$
calibrations, the matter of a metallicity-correction
required particular attention with the result that such
a correction is apparently not needed at the present level
of accuracy. Analysis of the color-magnitude and
period-color relations shows that globular cluster members
are under-luminous relative to the Galactic Disk contact systems 
mainly because of the smaller sizes and, consequently,
shorter orbital periods; the color-index
effect of the diminished blanketing  
is less important, especially for $(V-I)$. Among 
the Class-1 members (deviations in $M_V$ smaller than 0.5 mag.), 
the most common are Blue Straggler (BS) systems. The 
variability amplitudes for the BS systems show a significantly
different distribution from that for systems below the 
Turn Off Point (TOP): The BS systems in the sample have only small
amplitudes while the distribution for the systems below the
TOP is peculiar in containing only large amplitude
systems. This difference may be linked to the relatively
small number of the detected Main Sequence contact
systems below the TOP as resulting from 
an observational selection effect due to the rapidly increasing
measurement difficulties below the TOP. As
a consequence, efforts at determining the frequency of 
occurrence of the contact systems below the TOP have been
judged to be premature, but the frequency among the
BS stars could be moderately well established at
about $45 \pm 10$ BS stars per one contact BS binary.

\end{abstract}

\keywords{clusters: globular -- binaries: eclipsing -- blue stragglers}

\section{Introduction}
\label{intro}

Our thinking about close binary stars in globular clusters (GC's)
has undergone a tremendous change in the last decade.
Once such binaries were thought to be totally absent in the 
GC environment, while now they appear to be of great 
importance to the dynamical evolution of the clusters. 
A large volume of research on the dynamical effects 
of the binary systems on the cluster evolution, including 
important and complex inter-relations with 
the evaporation and tidal-striping effects,
has been summarized in several reviews starting with 
\cite{hut92}, with updates in 
Sections 9.5 and 9.6 of \cite{mh97} and in \cite{mcm98}.
Solar-type contact binaries (also known as W~UMa-type 
variable stars) contain the least amounts of
angular momentum that binary systems made of Main Sequence
components can have. They represent last stages
of the angular momentum loss (AML) evolution of
primordial binaries or are one of the
products of the dynamical inter-cluster interactions. Together
with the RR~Lyr-type pulsating stars on the horizontal branch
and SX~Phe-type pulsating stars among Blue Stragglers,
the W~UMa-type binaries are the most common type of
variables in the GC's. 

The number of W~UMa-type binaries in GC's has expanded 
in the recent four years from 24 (\cite{mat96a})
to the current number of 86. This paper attempts to integrate
and analyze the available data for these
variables in order to establish most essential
results and to guide further research. The analysis is
similar in its goals to the study of the Galactic Disk 
systems which
compared the W~UMa binaries in the galactic field -- as seen in 
the direction of the galactic Bulge in the OGLE survey --
with those in several old open clusters (\cite{open} = R98).
Searches for contact binaries in GC's are much more
difficult than in open clusters. They require -- on top of
a generous allocation of observing time permitting
variability detection and monitoring over several nights --
at least moderate-size telescopes located in sites with
excellent seeing. Cores of some clusters remain
too compact for photometry of individual stars even in 
perfect seeing; such clusters must be observed
with the Hubble Space Telescope for which continuous
monitoring of variability over long periods of time is
difficult to arrange. It is expected, however, that
new methods of analysis of difference images, such as 
developed by \cite{ala98} and \cite{ala99}, 
will result in substantial progress in detection
and analysis of variable stars in very dense fields
close to the cores; one of the first
applications to GC's (RR~Lyr stars in the core of 
M5) is by \cite{ole99}. 

Because of the action of two possibly mutually reinforcing
types of processes, of dynamical interactions and of 
the magnetic braking, rather than one -- as in the Galactic
Disk and in open clusters, where only the latter mechanism 
would be sufficiently effective --
one may expect high frequency of occurrence of the contact
systems in globular clusters, possibly even higher than
in the Disk. It is, in fact, surprisingly high
in the Disk: As shown in R98, the Disk systems 
show an increase in the frequency 
of occurrence among F--K dwarfs over time, in the accessible 
range of 0.7 -- 7 Gyr, reaching the spatial 
frequency as high as about one such binary (counted as
one object) per 80 -- 100 single stars in the 
Galactic Disk. Contact
systems with spectral types earlier than about 
middle A-type and orbital periods longer 
than 1.3 -- 1.5 days are less common with currently un-measurable
frequency of occurrence, as shown in \cite{ogle3}; this
long-period cutoff may be a function of the parent population because
the contact binaries with periods up to 2.5 -- 3 days are known
to exists in LMC (\cite{macho}). The high
frequency of contact binaries with a gradual increase with age
in old open clusters is consistent with the prolonged 
magnetic-wind braking AML process acting over a time scale of
some 1 -- 5 Gyr (R98) and producing relatively 
long-lived contact systems. 

Contrary to expectations based on the above reasoning,
the preliminary indications from 
individual studies that reported discoveries
of contact binaries in GC's (and are cited in this
paper) do not confirm the high frequency of
occurrence in these clusters. As several
authors of such papers already remarked
-- but bearing in mind the tremendous technical difficulties --
the frequency appears to be relatively 
low, at the level of a small fraction of a percent and some 
clusters do not seem to have W~UMa-type
binaries at all. An attempt to assess this matter                
is presented in this paper. It is argued that, at this
moment, we cannot really say much about the frequency
of occurrence of the contact systems on the Main
Sequence, below the Turn-Off Point; however, the 
frequency among the Blue Stragglers appears to be high,
some 2 to 3 times higher that among the stars of the 
Galactic Disk.

The current paper consists of a description of the 
sample of clusters with contact binaries in Section~\ref{clusters},
then of the sample itself in Section~\ref{members}. 
The observed metallicity effects affecting the absolute-magnitude
calibration (which is used to select the members) 
and affecting the observed properties of the systems 
are described in Sections~\ref{met-cal} and \ref{low-met}. 
The color-magnitude and the period-color relations are
discussed in Sections~\ref{col-mag} and \ref{per-col}.
Details concerning systems with EB-type light curves,
Blue Straggler contact systems and the frequency of
occurrence are given in Sections~\ref{EB} -- \ref{freq}.
Conclusions are stated in Section~\ref{concl}.

\section{The cluster sample}
\label{clusters}

The sample of GC's surveyed deep enough to include 
Main Sequence stars, below the Turn-Off Point (TOP), 
currently consists of 14 clusters. The sample is quasi random in
the sense that several authors contributed the data using their
own preferences, but that practically all data have 
come from ground-based telescopes. Dr.\ Ka{\l}u\.{z}ny and his
collaborators, who contributed most of the results, selected
primarily the nearest clusters, with moderately developed cores
(permitting photometry close to the centers) and avoiding 
those with small galactic latitudes, within $|b| < 10^\circ$.
The essential parameters characterizing the clusters
are given in Table~\ref{tab1}, with clusters arranged
according to the NGC number. To insure uniformity of these 
parameters, they have been taken from the database of \cite{har96}, 
version June 22, 1999, which is available at:
http://physun.physics.mcmaster.ca/Globular.html. In Table~\ref{tab1}
we give the following parameters: the galactic coordinates $l$, $b$
in degrees, the galacto-centric distance $R_{GC}$ in kpc, the reddening
$E_{B-V}$, the observed distance modulus $(m - M)_V$, the 
metallicity parameter $[Fe/H]$ and the concentration parameter $c$
(for collapsed cores, $c = 2.5$). 
The galacto-centric distances span a wide range $3.5 < R_{GC} < 18.5$
kpc while the metallicities occur within 
the representative range $-2.22 < [Fe/H] < -0.73$.
The sample is dominated by clusters with moderate and low concentration;
clusters with $c \rightarrow 2.5$ are under-represented. Eight
clusters have galactic latitude $|b| < 20^\circ$ indicating a possibility
of large interstellar extinction and heavy Milky Way stars 
contamination; for three clusters, NGC~4372, NGC~6121 and NGC~6441,
the reddening is large, $E_{B-V} > 0.3$. 

The cluster sample requires some comments: 
\begin{enumerate}
\item While Table~\ref{tab1} lists 14 clusters, 
the entry for NGC 5904 (M5) actually
reports a null result because all the systems suggested
as contact binaries by \cite{yr96} turned out to be spurious
detections, as has been shown by \cite{kal99}. The cluster has
been monitored for variability of its stars by several investigators
(see references to Table~\ref{tab1}).
\item NGC 4372 is a very important cluster because
of its low metallicity and a large number of systems discovered
in its direction. Unfortunately, because of its low
galactic altitude of $b = -10^\circ$, 
it is has a large and patchy reddening. The discovery paper
by \cite{kk93} gives, in its Table~3, 
the observed data in $(B-V)$ and $(V-I)$,
but the patchiness-corrected data are given 
only for the former color index.
The corrections have been calculated for $(V-I)$ assuming the
relation $\Delta E_{V-I} = 1.24 \, \Delta E_{B-V} = 
1.24 \, [(B-V)_c - (B-V)]$. 
The same slope is used throughout this paper in deriving the 
values of $E_{V-I}$ from $E_{B-V}$. 
On top of the patchy reddening, the values of
the mean reddening and of the 
distance modulus are very uncertain for this
cluster. While Ka{\l}u\.{z}ny \& Krzeminski assume $E_{B-V} = 0.48$ and 
$(m-M)_V = 14.8$, the Harris database (Table~\ref{tab1} here)
quotes $E_{B-V} = 0.39$ and
$(m-M)_V = 15.01$. For consistency, the Harris set has been used here,
but -- as we discuss in the next Section --
this choice makes a very important difference in ascertaining
membership of systems detected in this cluster.
\item The color-magnitude of the core of NGC~6752 has been studied
with the Hubble Space Telescope (\cite{rb97}). It indicates a 
relatively high frequency of binary stars at a level of $15 - 38$
percent in the inner core, but below 16 percent beyond the core.
The relative frequency must be therefore a strong function
of the radial distance from the cluster center. Unfortunately, 
searches for variable stars in the clusters analyzed in this
paper are not uniform in this respect: Some clusters were
observed with the inclusion of the cores, some were observed
only at some distance from the center
where crowding was assumed to be tolerable. This casts a large
uncertainty on any considerations involving numbers of contact
systems and their frequency of occurrence relative to other
stars.  
\end{enumerate}

\section{Cluster members}
\label{members}

All contact binary systems discovered in globular clusters are listed
in Table~\ref{tab2}. The Fourier-analysis 
of the light curves was not used to verify the W~UMa-type
or the EW shape of the light-curves
(as in the selection of the OGLE sample in \cite{ogle1}) 
because of the partial coverage of 
some of the light curves which could produce incorrect values
of the Fourier coefficients; the general appearance of
the light curve and the original classification by the
discoverers were the only criteria used here. 
Contact systems with
unequally deep eclipses (EB-type light curves) have been retained;
they are marked as such in Table~\ref{tab2}. Systems having 
light curves suggesting detached components (EA type) are
not considered here.

The empty entries in Table~\ref{tab2} are due to the fact
that most of the photometric searches have been done 
either in $B$ and $V$ band-passes or $V$ and $I_C$ band-passes 
(the subscript indicating that the $I$ 
band is of the Cousins system will not be used from now on). 
Thus, the available data split in two sets, 
forcing us to discuss all relations and all properties 
in parallel in $V$ and $(B-V)$ and in $V$ and $(V-I)$. We will call
these the $BV$-set and the $VI$-set. 
Only one cluster was observed in all three bands,
NGC 4372 (\cite{kk93}) permitting consideration of two photometric 
indices, $(B-V)$ and $(V-I)$, for 8 contact systems in this cluster.
As luck would have it, this is the cluster which is the
most heavily reddened with a highly patchy and uncertain extinction.

Table~\ref{tab2} lists the variables with the names and
designations as assigned in the
discovery papers. For each system, we 
give the orbital period in days, then the
maximum brightness $V$, $(B-V)$ and/or 
$(V-I)$ and the total variability 
amplitude $A_V$. The last columns give our results on the membership
to the parent clusters and the variability/membership type (see below).
Assignment of the membership has been performed by comparison
of ``observed'' absolute magnitudes, derived from the distance modulus: 
$M_V^{obs} = V - (m-M)_V$, with those derived from the calibrations,
$M_V^{BV} = M_V(\log P, B-V)$ or $M_V^{VI} = M_V(\log P, V-I)$, as
described in more detail below. A simple
criterion for membership was used calling systems with the deviations 
in $\Delta M_V = M_V^{obs} - M_V^{cal}$ 
smaller than 0.5 mag the Class~1 members and those
with $0.5 < |\Delta M_V| < 1.0$ the Class~2 members (here
$M_V^{cal}$ takes the meaning of $M_V^{BV}$ or $M_V^{VI}$
depending which one has been available). Deviations
larger than one magnitude were assumed to signify that the binary is
not a member but a foreground or background projection. 

The calibrations used in this paper have been:
\begin{eqnarray}
M_V^{BV} = -4.44 \, \log P + 3.02 \, (B-V)_0 + 0.12 \label{eq1} \\
M_V^{VI} = -4.43 \, \log P + 3.63 \, (V-I)_0 - 0.31 \label{eq2}
\end{eqnarray}
The $M_V^{BV}$ calibration is based on the Hipparcos data
(\cite{rd97}) while the one for $M_V^{VI}$ was
developed and served well for the analysis of all the OGLE data
(\cite{ogle1}, \cite{ogle2}, \cite{ogle3}, \cite{open}). 
No allowance for lowered metallicity
of GC's in the calibration has been made. 
We discuss this important issue
in the next Section~\ref{met-cal}.

The deviations $\Delta M_V$ are shown in graphical form in
Figure~\ref{fig1} versus the orbital period and color index,
for both color-index sets. While many systems do fall within
the band of $\Delta M_V$ close to zero, a large fraction
of systems are actually foreground projections onto the
fields of the observed clusters. This fully agrees with the high
frequency of the W~UMa-type binaries in the Galactic Disk. The data
based on the $(V-I)$ index show a somewhat better consistency, with
smaller scatter in $\Delta M_V$ and with 
a better definition of the group of the Class-1 members. This may be due
to a weaker dependence of this index on the interstellar reddening 
and/or on the metallicity, but may possibly be simply due 
to lesser photometric difficulties of observing red stars in the
$V$ and $I$ bands than in the $B$ and $V$ bands. As pointed out
to the author by Dr.\ Ka{\l}u\.{z}ny (private communication), 
selection of the best photometric bands is not an easy matter so
that $(V-I)$ may not always be preferable to $(B-V)$:
The upper MS is better observable in $B$ than in $I$ because the
numerous red dwarfs from the lower MS produce a strong background
in $I$; the $I$ band is also inconvenient for red stars because
Asymptotic Giant Branch stars are usually strongly over-exposed
and prevent photometry of other stars close to cluster centers.

The two penultimate columns of Table~\ref{tab1}
give the number of contact systems in the GC's. $n_{det}$ is the total
number of systems in a cluster, while $n_{C1}$ and $n_{C2}$ are
the numbers of Class-1 and Class-2 members, respectively.
$n_{BS}$ is the number of Blue Stragglers (of both
classes). $N_{BS}$ are very approximate estimates
of Blue Stragglers which were monitored for variability
in the GC's. We will discuss these data later, in Sections~\ref{ampl}
and \ref{freq}. 
The approximate
locations of the color indices at the Turn-Off Point are
marked by vertical broken lines in Figure~\ref{fig1}. Note that
practically all systems to the blue of these lines (i.e. the
Blue Stragglers) are members of the clusters. 

The cluster-member selection process described above assumes that the
calibration formulae, Eq.~\ref{eq1} and Eq.~\ref{eq2} 
are applicable to contact systems in globular clusters. This is
verified by assuming that most typical systems which have been
detected in a given direction 
are actually genuine members of the clusters.
In effect, we require that the calibrations 
reproduce the modal (most probable) values of the deviations
$\Delta M_V$. The histograms of the deviations 
are shown in Figure~\ref{fig2}. Disregarding a complication
of the metallicity-dependence of $M_V^{cal}$ which will be 
discussed in the next section, we can observe the following
facts:
\begin{enumerate}
\item There are 35 Class-1 systems in the sample, that is
roughly 1/2 of the total. 
\item The number of the cluster members
would increase by four if we add Class-2
systems which have blue color indices and short periods
characteristic for Population~II Blue Straggler systems
(Section~\ref{ampl}). 
\item We have no good argument to claim that any of the
eight remaining Class-2 systems is a cluster member. 
\item There are 21 definite non-members in the sample, 
that is about 1/3 of the total number. All but one
are foreground Disk systems. 
\item The only obvious {\it background\/} system is 
V6 in NGC 6752. The $\Delta M_V$ deviation of about 2.5 mag.\
suggests that the system is some 3 times further away than the cluster,
at a distant periphery of the Galaxy at the distance of some 15 kpc.
\item As discussed in \cite{kal98a}, there exists an ambiguity with the
orbital period for the only 
contact system detected in M3, V238. It appears
that neither of the acceptable periods places the system in the cluster.
The two possibilities are joined by a dotted line in Figure~\ref{fig1}.
Dr.\ Ka{\l}u\.{z}ny (private communication)
suspects -- from analysis of the individual
CCD images -- that blending with a red giant is the cause
of photometric difficulties with V238.   
\end{enumerate}

\section{Metallicity effects in the $M_V$ calibrations}
\label{met-cal}

The cluster-selection process requires a clear conceptual separation
of the effects of lowered metallicity 
on the simplified $M_V$ calibrations given by Equations
\ref{eq1} and \ref{eq2} (they are 
called here $M_V^{cal}$ or specified as $M_V^{BV}$ or $M_V^{VI}$) 
which we use for selecting the cluster members
from the effects genuinely influencing the observed binary 
properties, such as the
absolute magnitudes $M_V^{obs}$ (derived from cluster moduli), 
the de-reddened color indices or the orbital periods. Here
we concentrate on the effects affecting $M_V^{cal}$
and solely from the observational point of view, without going 
into the details of how genuine properties are affected;
this will be discussed in the next Section~\ref{low-met}. 
We note that once we
select a sample of the cluster members, we will discuss only the
directly observed quantities; the values
of $M_V^{cal}$ will not be used from this point at all.

In \cite{cal2} arguments have been presented that the 
$M_V^{cal} = M_V(\log P, {\rm color} )$ 
calibrations require small, but significant
corrections for metallicity variations. The corrections would reflect the
blue colors of Population~II stars which have low 
atmospheric blanketing due to weak spectral lines of metals. 
It was argued that while the period term in
Equations \ref{eq1} and \ref{eq2} would account for
any differences in the system size, the color-index term
would require a correction as a proxy of the effective
temperature. The data and the available 
calibrations were very preliminary at that time;  
the current material is much richer so that the need for the
corrections can be re-evaluated. This matter is very closely
linked to the selection of the cluster members, so that we must
explain the details of this process.

Figure~\ref{fig3} shows the same deviations 
$\Delta M_V$ as in Figure~\ref{fig1}, but this time plotted 
versus $[Fe/H]$ for the clusters. The slanting lines
give the metallicity corrections $\delta M_V^{BV} = -0.3 \, [Fe/H]$
and $\delta M_V^{VI} = -0.12 \, [Fe/H]$, as suggested in \cite{cal2}.
To visualize how the corrections would impact the membership
selection process, the $\Delta M_V$ deviations --
with the metallicity corrections applied -- are shown as broken-line
histograms in Figure~\ref{fig2}. 

Figures~\ref{fig2} and \ref{fig3}
contain the only available information on the need of the
corrections and how their use would influence the selection of the
cluster members. 
The situation is relatively simple for the $VI$ set: 
Because of the well-known low sensitivity of this color index to
metallicity variations, the current 
data do not forcefully suggest or reject the need
for the metallicity corrections, although the case of
no corrections seems to be slightly preferable; for simplicity,
we assume that they are not needed. The data for
the $BV$-set show a large scatter in the band of the 
expected members and its is difficult to decide if the
corrections are really needed.
The crucial clusters for resolution of the problem 
for the $BV$ set are the two clusters with the 
lowest metallicities $[Fe/H] < -2$, NGC 4372 and NGC 5466. 

As remarked before, NGC 4372 
is seen at low galactic latitude through a patchy interstellar
extinction. However, the cluster 
was observed in $BVI$ so that two sets of color indices
are available and a consistency check is in principle
available. Assuming the data as in Harris' database
(as given in Table~\ref{tab1}), 
the $M_V^{VI}$ calibration indicates that {\it none of
the contact systems can be classified as Class-1 member\/} while 
the $M_V^{BV}$ calibration suggests that one system,
V4, is a Class~1 member. However, if we follow
the assumptions of \cite{kk93}, $E_{B-V} = 0.48$ and $(m-M)_V = 14.8$,
then large shifts in $\Delta M_V$
by 0.49 and 0.61 mag.\ for both color-index sets
occur: While $M_V^{obs}$ become fainter by $+0.21$ mag., 
the predicted $M_V^{BV}$ and $M_V^{VI}$ become brighter by
$-0.28$ and $-0.40$ mag. As the result, 
all data points for this cluster slide down 
in Figure~\ref{fig3} by amounts shown there by arrows. While
for the Harris data only the system V4 would be a Class-1
member in the $BV$ set (without a confirming evidence from the
$VI$ set), now the systems V4, V16 and V22 would be Class-1 members
for the $BV$ set and V5, V16 and V22
would be Class-1 members for the $VI$ set. One gains
then in consistency between both sets for V16 and V22, but
the matter of membership 
remains unclear for V4 and V5. It is obvious that
only three or four systems among 
the eight might be members of
NGC~4372, but we cannot be absolutely sure which 
ones. In this situation, 
we have taken a conservative approach and conclude
that the data for NGC 4372 are too uncertain to be sure
of the membership of the systems in this cluster; this
cluster cannot tell us much about the need for a metallicity 
correction in the calibration. We note that judging by its
period-color combination, the system V22
is a genuine Blue Straggler and irrespectively how big
is its deviation $\Delta M_V$, it is almost certainly a member
of NGC~4372 (see Section~\ref{ampl}).

NGC 5466, with even more extreme metallicity than that of NGC~4372
of $[Fe/H] = -2.22$, follows the Hipparcos calibration for 
the Disk stars (\cite{rd97}), without any metallicity
correction, very well. The two contact
binaries must be genuine members simply because at 
the galactic latitude of $b = +74^\circ$ chances of
having Disk population stars within the cluster are practically
equal to zero. Also, both stars are exceptionally blue, belonging
to the Blue Straggler group of contact binaries 
(see Section~\ref{ampl}); equally blue
systems very rarely occur in the Galactic Disk.

Guided mostly by the case of NGC 5466 and by the resulting
simplicity of the assumption,
we conclude that the $M_V$ calibrations established
for $[Fe/H] = 0$ apparently work well for low values
of metallicity and that the $[Fe/H]$-corrections in the expressions
for $M_V^{cal}$ are apparently not needed. 
As we will see in the next sections,
the color indices of the low-metallicity systems
are definitely much bluer than those for Disk systems, so that
-- without any corrections -- 
one would expect values of $M_V^{cal}$ indicating
artificially higher luminosities. Since this is not observed,
some other metallicity-dependent factor must provide a compensating
effect through the period-term (this may be for example
the un-accounted for influence of the mass-ratio).
Concerning the previous, apparently erroneous result in \cite{cal2}, 
we remark here that the correction of $- 0.3 \times [Fe/H]$
for the $(B-V)$-based calibration was suggested for an old version
of the $M_V^{BV}$ calibration which has been
superseded by the much better Hipparcos calibration. The
corresponding correction for the $(V-I)$-calibration
of $-0.12 \times [Fe/H]$
was suggested for consistency with the one for $(B-V)$, but it was
always recognized that it was smaller than the measurement and
definition uncertainty of the calibration itself.

\section{Expected effects of low metallicity}
\label{low-met}

In summarizing the main effects of lowered metallicity, we discuss
two effects which manifest themselves very differently: of the blue
atmospheres (for the same effective temperature) and of smaller
stellar sizes. 

The decreased metallicity influences the atmospheric 
structure in that metal lines produce less blanketing so that stars 
become bluer. Only this effect
was discussed in \cite{cal2}, together with its possible influence
onto the $M_V$ calibrations. For consistency with the previous results, 
we use the same relations between the color-index changes and $[Fe/H]$
as evaluated from the models of \cite{bk92}. The expected changes in the
color indices $(B-V)$ and $(V-I)$ are given for a MS star atmosphere
at $T_{eff} = 5000$~K in Table~\ref{tab3}.
Obviously, several qualifications
may be in order here: Single-star model-atmosphere results may be
in-applicable to magnetically active contact binaries and the color
index data have been calculated for spherical stars whereas 
strong and variable limb darkening effects are always present in 
contact binaries. 

The low-metallicity systems are expected to have
smaller dimensions and -- for
the same contact geometry as for Galactic Disk systems -- should have
shorter periods than Population~I systems. 
In the currently sole study on properties of
Population~II contact systems binaries by \cite{web79}, the stress
was on the influence of the prolonged
angular-momentum loss for such old
objects. The author, however, remarked about the smaller sizes of
such systems relative to Population~I systems, but did not 
discuss this point any further.
Some guidance on the expected effect can be found in the mass-radius
relation and its dependence on the stellar metallicity. For a fixed
mass, Kepler's law enforces the proportionality 
$\log P \propto 3/2\, \log A$. If the geometry of contact is 
the same irrespective of the chemical composition so that
the {\it relative\/} sizes of components are 
independent of the metallicity, $r_i = R_i/A$, then
smaller stellar sizes $R$ should lead to smaller $A$ and
to shorter orbital periods. Again, a qualification
may be in order: The period changes
may depend on metallicity in a much more complex way than just through
simple radius scaling because the internal structure of Population~II
contact binaries does not have to be identical to that of Population~I
systems. 

The mass-radius relation for low-metallicity stars is currently a subject
of very lively discussions, stimulated by the distance determinations for
subdwarfs, as provided by the Hipparcos mission. One of the most recent
theoretical investigations of stellar models with varying abundances
is by \cite{cas99}. The absolute magnitude calibration for
stars with $T_{eff} = 5000$~K (Eq.~(1) in this paper)
can be re-written into a metallicity-radius dependence:
$\Delta \log R = 0.158\,[Fe/H] + 0.035\,[Fe/H]^2 - 0.485 \Delta Y$;
with the solar abundance assumed to be $Z_\odot = 0.0169$
and $[Fe/H] = \log Z - \log Z_\odot$. The size of the correction for
helium abundance changes, $\Delta Y$, which accompanies the metallicity
changes as the stellar 
population ages, is a difficult matter. The authors
discuss the large uncertainty in the ratio $C = \Delta Y/\Delta Z$ which
is currently very poorly known; some observational results 
suggest $C = 3 \pm 2$ while the authors consider $C \simeq 5 - 6$. 
The metallicity-radius relation is shown in Figure~\ref{fig4} for three
values of $C$. 

With the expected values of 
$\Delta \log P \propto 3/2 \times \Delta \log R$ 
one can compute the expected variations of the absolute magnitude.
We have a choice here: We can simply assume that
the luminosity will scale with the square of the radius
or we can use the $\log P$ terms in the Equations \ref{eq1} and \ref{eq2}.
The former choice would suggest: 
$\Delta M_V = -5 \times \Delta \log R$, while
the latter choice gives a steeper dependence:
$\Delta M_V = -4.44 \times 3/2 \times \Delta \log R$
(the period-term coefficients are 
basically identical for both color-index
sets). In what follows, we will use the second expressions 
for consistency with the adopted expressions for $M_V^{cal}$,
observing that the steepness of the period term (which probably hides
many unaccounted period-dependent effects) 
may actually explain the unexpected absence of the $[Fe/H]$ term
in the equations giving $M_V^{cal}$.

\section{Color-magnitude diagram}
\label{col-mag}

The color-magnitude diagrams for the sample of the GC members are
shown in Figure~\ref{fig5}. The data plotted are the
observed $M_V^{obs}$ and the de-reddened color indices so that
uncertainties with the metallicity corrections in $M_V^{cal}$ 
do not enter directly into this figure (but only through
the selection of cluster members, via the deviations $\Delta M_V$). 
The Class-1 members, which we consider genuine members of the
clusters, are marked by filled circles and the range of metallicity
($[Fe/H]$ smaller or larger than $-1.5$) is shown by the
size of the symbol. The figure also contains the theoretical
isochrones computed by the Padova group (\cite{ber94}) for 14 Gyr and for
three values of metallicity $[Fe/H] = -1.66$, $-1.26$ and $-0.66$.
In addition, the data for members of old open cluster members are shown
by small x-symbols, following the results in R98.

The most striking feature of the color-magnitude diagrams for
Population~II contact binaries is the shift of the 
contact-binary sequence by about one magnitude
below that for Galactic Disk systems. A similar shift is 
well known for single subdwarfs, but it is seen here 
for contact binaries for the first time. 
The sequence is relatively well defined and
extends uniformly on both sides of the Turn-Off Point (TOP), similarly as
in old open clusters such as Cr~261 (R98).

A group of Blue Stragglers (BS) in the low-metallicity
clusters, to the blue of the TOP is the most conspicuous group 
of the contact binaries in the color-magnitude diagrams in
Figure~\ref{fig5}. 
A few Class-2 systems also have very blue color indices 
of the BS stars; since such systems practically
do not occur among Disk contact systems we think that they 
are not foreground projections, but genuine
cluster members with
the $\Delta M_V$ deviations larger than 0.5 mag. These systems
are in the $BV$ set: V22 in NGC~4372 and V8 in NGC~6752;
in the $VI$ set: V22 in NGC~4372 and V65 in NGC~5139. Thus, 
V22 in the controversial cluster NGC~4372,  
with its Blue Stragglier characteristics, is  
intrinsically the brightest system in the current GC sample
with $M_V^{obs} = 1.83$, $(B-V)_0=0.24$ and $(V-I)_0=0.30$.

\section{Period-color diagram}
\label{per-col}

The period-color diagram is, to some degree, a more natural and precise
way to display properties of contact binaries than the color-magnitude
diagram. This is because one quantity, the orbital period,
is known basically without error, at least when compared
with photometric errors. Only one photometric quantity --
the color index -- then enters into the picture. The color index
can still be affected
by several factors, in addition to the measurement errors, the
most severe being the uncertainty in the reddening correction, but
the period-color relation is not affected by errors in the distance
modulus.

Figure~\ref{fig6} shows the two period-color diagrams for the
GC sample, with the same symbols as in Figure~\ref{fig5}. The reader
is suggested to view both figures simultaneously and to note the common
features. The figure contains the Short-Period Blue-Envelopes (SPBE)
for Disk systems, shown as continuous lines for both color-index sets
(\cite{open}, \cite{ogle1}). Their shapes are given
by: $(B-V)_{SPBE} = 0.04 \,P^{-2.25}$ and 
$(V-I)_{SPBE} = 0.053 \,P^{-2.1}$ (the orbital period $P$ is in days). 
While the numerical values in these definitions do not have any 
physical meaning, the curves are important 
because they delineate location
of the least-evolved contact systems in the Disk Population sample.
As many theoretical investigations indicated, contact systems
would normally evolve away from the main Sequence, toward larger
stellar dimensions, longer orbital periods and cooler atmospheres.
The interstellar reddening also increases the color index. Thus,
the SPBE has a meaning of a Zero-Age Main Sequence for the Disk
population systems.

The above considerations should also apply to Population~II contact
systems except that, as pointed out by Webbink, the angular-momentum
loss due to gravitation radiation emission may win over
the long time scales involved and eventually lead to shortening
of the orbital
periods. As we see in Figure~\ref{fig6}, the GC systems indeed have
periods shorter than those for Disk systems, but this is expected
irrespective of whatever mechanism of forming them is involved: 
As discussed in Section~\ref{low-met}, to be a contact system,
a low-metallicity binary must have a short period because its
components are small. Judging by the size and direction of the
metallicity corrections (shown by arrows), the relatively 
larger effect is observed in $\Delta \log P$ than in the 
color-index shifts $\Delta(B-V)$ and $\Delta(V-I)$.
Although the two changes can partly compensate 
each other in their control of the absolute magnitude, 
the compensation is not exact because the arrows in Figure~\ref{fig6}
are not parallel to the lines of 
constant $M_V$. This was visible in the color-magnitude
diagrams in Figure~\ref{fig5}, where the GC systems were obviously
fainter than the Disk systems. It is striking how different
Population~II contact systems are from the Galactic Disk systems,
but also how similar they are within their group.
Apparently, a difference in metallicity from the solar
$[Fe/H] \simeq 0$
to $-0.7$ or so produces a relatively larger change in 
their properties than a further change to $-2.2$ observed
for the most extreme-metallicity systems. Since most changes
are due to the change in the component dimensions, this would argue
for a relatively small value of the currently poorly-known coefficient 
$C = \Delta Y/\Delta Z$ (see Figure~\ref{fig4} in Section~\ref{low-met}).

\section{EB-type systems}
\label{EB}

The sample of binary stars considered in this paper contains six
systems with EB-type light curves. Such light curves are characterized
by unequally deep eclipses, but with strong variations between
minima suggesting possibility of a physical contact. The OGLE sample
(\cite{ogle2}) contained only 2 systems of this type among 98
systems in a volume-limited sample of contact binaries selected
using a Fourier light-curve shape filter. In this paper, we do
not use this filter because it is sensitive to the phase coverage
and tends to be too discriminatory. Thus, we assume that the
six systems are related to contact systems. In fact, these may
be various forms of semi-detached binaries in the pre-contact
or broken-contact stages, with either the more
massive, hotter or less-massive, cooler components filling their
Roche lobes (\cite{egg96}). The latter cases appear to be less common,
but a good case of a very short period Algol has been recently
identified in W~Crv (\cite{rl99}).

Three among the six systems appear in the $BV$ set and three in the
$VI$ set. Inspection of Figure~\ref{fig1} and Table~\ref{tab2}
shows that all but one are Class-1 cluster members; the one with 
a slightly larger $\Delta M_V$ is a Class-2 member. Thus, they all
follow the absolute-magnitude calibrations for 
normal contact binaries and seem to be genuine cluster members.
This relatively high frequency of occurrence among contact systems,
of 6 among 35 (or 39 if 4 Class-2 Blue Stragglers are added)
is unexplained and interesting. We note that two among the six systems
are Blue Stragglers.

\section{Amplitude distribution and the Blue Stragglers}
\label{ampl}

It has been noted in Section~\ref{col-mag}
that the contact system sequence continues
without any obvious changes in the
period, color or absolute magnitude properties
across the Turn-Off Point,
into the Blue Stragglers domain. Now we will address the
only property which is available to characterize the light
curves, the variability amplitude and its distribution. 

The Blue Stragglers (BS) are an important group of stars in old stellar
clusters. It is now recognized that they must form through
binary evolution processes, although it appears that there are actually
many such processes and it is not easy to find out which one
occurs most commonly (\cite{leo96}, \cite{mat96b}). The BS formation
and evolution is such a large and active area that special
meetings have been devoted to it (\cite{saf93}) and very
active research continues. 
Contact BS are relatively easy to
identify for several reasons: (1)~They are bluer than typical
galactic Disk systems which start appearing at $(B-V)_0 < 0.3$
(R98), 
(2)~Their photometry is relatively less difficult than for
the Main Sequence stars 
because they are photometrically well above the level where
-- usually formidable -- crowding problems for the MS stars
set in. The question
is: Are they in any other way different from 
the ``normal'' Main-Sequence contact systems in the GC's? One such
property can be amplitudes of light variations. 
Although the amplitude statistics 
involves a convolution of the distribution of the mass-ratios
with the distribution of orbital inclinations,
as was discussed in \cite{ogle2}, 
lack of large amplitudes must mean that the large 
($q \rightarrow 1$, $q = M_2/M_1 \le 1$) 
mass-ratios do not occur: When components differ
in sizes, only small amplitudes are possible.

Figure~\ref{fig7} shows the observed amplitudes for
the GC members plotted in relation to the intrinsic color 
indices. For the color index values, we assume
that systems with $(B-V)_0 < 0.4$ or $(V-I)_0 < 0.55$
are Blue Stragglers. The amplitudes do show a 
change at the Turn-Off Point, but this change is well defined
only for the $VI$ set: The large amplitudes are observed only for the
systems to the red of the TOP, that is for the genuine MS systems.
The two-sided Kolmogorov--Smirnov tests comparing distribution
on both sides of the TOP's, limited to Class-1 systems, 
indicates that the difference in the distributions is
not significant for the $BV$ set, but the probability
of a random chance producing the observed
difference for the $VI$ set is only $1.1 \times 10^{-5}$. The
significance changes only slightly when Class-2 members
are added with the probabilities of a random result of 
0.05 for the $BV$ set and $0.71 \times 10^{-5}$ for the $VI$ set. 
The change in the observed amplitude distribution at the TOP
is therefore very highly significant for the $VI$ set,
but insignificant for the $BV$ set. At this moment, we have no
explanation for this difference 
between the color-index sets except that -- as pointed
several times in this paper -- it may be related
to the photometric difficulties in the $B$-band for red stars
and to the larger and more uncertain reddening corrections 
in the $(B-V)$ color index.

On the basis of the above numbers one
would be tempted to claim that, indeed, the BS's have smaller
amplitudes than the Main-Sequence contact systems. However, a caution
is in order here: {\it It is very important that
we do not see small amplitudes among the
Main Sequence systems below the TOP, 
only the large ones\/}. They are obviously missing there
because for random orbital inclinations 
systems showing small amplitudes
should be always more common than systems showing 
large amplitudes. The histograms in the upper 
parts of the panels of Figure~\ref{fig7} contain the expected
amplitude distributions calculated in \cite{ogle2} for two
assumptions of the mass-ratio distribution, a flat $Q(q)$
and $Q(q) = (1-q)$. The data for the Disk systems in the OGLE sample
suggested that the distribution may be actually even more strongly
decaying with $q \rightarrow 1$ than the one $\propto (1-q)$, 
because large amplitudes are exceedingly rare in an unbiased sample
(this is very much unlike the sky-field sample).
In the case of the present GC sample, a comparison of the
theoretical distributions with the observed ones suggests that
the observed data may be severely modified by strong
detection selection effects for systems below the TOP.
While almost all systems in the BS group appear to be detected
(there is only a small depression in the distribution
for $Ampl_V < 0.2$), the
MS systems appear to be entirely missed for $Ampl_V <0.3$
due to the difficulties with accurate photometry in conditions where 
measurement errors and crowding problems rapidly increase with
the apparent magnitude.

\section{Frequency of W~UMa-type systems in globular clusters}
\label{freq}
 
The discovery selection effect
described in the previous section may lead to an under-estimate 
of the number of the contact Main Sequence (below TOP) 
systems by a factor of the order of about 5 to 10.
The previous results for the sky-field sample should be recalled
here as a sobering experience: 
For several years the frequency of contact binaries of one per
one thousand MS stars has been considered  
a well-established, ``textbook-level''
fact, in spite of the warnings that the sky sample 
of known contact binaries contained
only large-amplitude systems (see Figure~2 in \cite{kr94}). 
Only the systematic characteristics of
the OGLE sample (R98) have shown that the 
apparent\footnote{We distinguish the apparent frequency,
which is uncorrected for missed low-inclination systems,
from the spatial frequency which is about 1.5 to 2 times
higher. The correction factor depends on the mass-ratio
distribution (\cite{ogle2}).} frequency is about ten
times higher, reaching about 100 -- 130 normal F--K dwarfs per
one contact system. 
A very similar situation re-emerges here,
except that this time we can directly suspect -- from the
difference in the amplitude distributions on both sides of the
TOP -- that the discovery selection effects are
more severe below the TOP. We cannot correct for these
selection effects because they must be different for each cluster
and are very difficult to quantify. 

In this situation, it has been
felt prudent to abandon attempts of determining the frequency
below the TOP and concentrate on the frequency 
data for the BS systems. This is in turn complicated by lack of
information in discovery papers on the number of Blue Stragglers
which were monitored for variability. A simple, but potentially
risky assumption has been made at that point 
that the numbers of the BS's shown in diagrams in the discovery 
papers are equal to the numbers of systems which
were actually monitored. Frequently stars without measurable
color indices (and not shown in color-magnitude diagrams)
are monitored for variability, but one may hope
that this would not happen in the BS region. Thus,
approximate numbers of the Blue Stragglers, $N_{BS}$,
have been estimated by
simply counting of data points on the color-magnitude diagrams.
It is stressed that these are very approximate estimates; for
example, for $\omega$ Cen the available estimate for the 
fields D--F (\cite{kal97b}) was multiplied by two to obtain
the total number in all observed fields. For 47~Tuc (\cite{kal98b})
unpublished data have been supplied by Dr.\ Ka{\l}u\.{z}ny.
The estimates are given in Table~\ref{tab1} and should be compared
with the number of the contact BS systems, $n_{BS}$. In doing so
we immediately see the problem of numerically very small values
for $n_{BS}$ which are obviously subject to 
relatively large Poissonian fluctuations.
Thus, we can see that we
are still very far from being able to correlate the
frequency of contact BS systems with cluster properties such as
the metallicity index, $[Fe/H]$, or the concentration parameter, $c$.
In this situation another disputable step was made in assuming
that the average frequency is the same for all GC's. This can be
derived by simply summing $N_{BS}$ and $n_{BS}$ 
for all clusters and taking their ratio.
The result is that 20 contact binaries are observed among
about 900 Blue Stragglers. Thus the average inverse
frequency is $f_{BS} = 45 \pm 10$ normal BS stars per one contact system. 
This frequency is the apparent one, that is it 
applies to systems which can be discovered, without any corrections
for systems missed because of the low orbital inclinations.
Since we do not know the mass-ratio distribution for the
contact BS systems, we cannot correct for these
missed systems to evaluate the true spatial frequency.

The inverse apparent frequency of
$f_{BS}=45$ is significantly different from 
the one observed for the Disk stars, which is approximately
$f_{Disk} \simeq 100 - 130$. Thus, the Blue Straggler 
population of the globular clusters contains some
$2-3$ times more contact binaries than the Old Disk stars. We note
also that the inverse frequency at the level of $45 \pm 10$
is in perfect accord with lack of detections in clusters poor
in Blue Stragglers where their absence can be
simply explained by the Poisson fluctuations.
 
The high frequency of contact binaries among Blue Stragglers
is visible in the color index and period distributions
shown in Figure~\ref{fig8}. These distributions are obviously
far from being rigorous in the statistical sense, yet they do
show interesting trends. When compared with the Disk systems (R98),
the GC contact binaries occupy only the blue ends of the
$(B-V)_0$ and $(V-I)_0$ distributions. Note, however, that the blue
end points are the approximately same for the GC and Disk sample
distributions indicating that the Disk sample may contain
an admixture of low-metallicity objects similar to those in
globular clusters. While the red systems are mostly 
likely under-represented in the GC sample 
because of the selection biases against
faint, red systems, we see a definite lack of long-period 
systems which are intrinsically the brightest and should 
be easily detected. Of course temporal-window biases for 
the GC sample may have contributed here (in the sense that monitoring
programs were by necessity short), 
especially when compared with the excellent
data for the 5 kpc OGLE sample (R98) which defines
the long-period part of the period distribution in 
Figure~\ref{fig8}. 

\section{Conclusions}
\label{concl}

Although at least 1/3 among 86
systems presumably located 
in the analyzed globular clusters are foreground
projections from the Disk which must be carefully
weeded out, we can state with confidence that
contact binaries in globular clusters are definitely different
than the very common Disk population systems. The main
feature are their short orbital periods resulting from
small dimensions of components. This is seen not only in
the distribution of the orbital periods, but also in low
luminosities. Thus, the downward shift of the contact binary
subdwarf sequence below that for the Disk systems is primarily
due to the reduced-dimension effect, not to the blue shift
caused by the reduced blanketing. The long-period
systems are intrinsically more luminous and easier to
discover so that their absence is highly significant. 

While the metallicity effects  
are clearly seen in the properties of the Population~II
contact binaries, to our surprise, they are not
visible in the $M_V^{cal}=M_V^{cal}(\log P, color)$ 
calibrations.  
More exactly, they do not manifest themselves in the calibration
based on the $(V-I)$ color index; the data utilizing
the $(B-V)$ color index are too poor to be sure that
the calibration does not need a metallicity term. For 
simplicity, we assumed that both calibrations (which are
used only to select the systems, not to analyze them)
do not require any $[Fe/H]$ dependent terms. It is
recommended that the $VI$ bandpass filter combination be used
in the future because the $BV$ data may indeed
show some metallicity dependence, but -- more
importantly -- are more susceptible to photometric
errors for red stars and to reddening correction
uncertainties.

Very little can be said about contact binaries located
on the Main Sequence, below the Turn-Off Point (TOP). 
Severe discovery selection effects are suspected from the
peculiar distribution of amplitudes with missing small
amplitudes. In contrast, 20 Blue Straggler contact binaries
known at this time give a reasonable estimate 
of their frequency of occurrence of one such
system (counted as one object)
per $45 \pm 10$ Blue Stragglers. This 
reciprocal apparent frequency is about 2--3 times higher
than for the Disk systems among the normal F--K stars. 
It is entirely possible that the same mechanism
which produces a continuous sequence of contact binaries
across the TOP in old open clusters simply had more
time to produce more contact systems in globular
clusters.

\acknowledgements
The author would like to express his indebtedness
to the authors of the original papers on individual
globular clusters. Special thanks are due to Dr.\ Janusz
Ka{\l}u\.{z}ny for his particular contribution to the field
and for his enthusiastic help in various stages of
this work and for detailed suggestions. Thanks are due to
Dr.\ Bohdan Paczynski for his useful comments on the
draft of the manuscript.

\newpage

\noindent
Captions to figures:

\figcaption[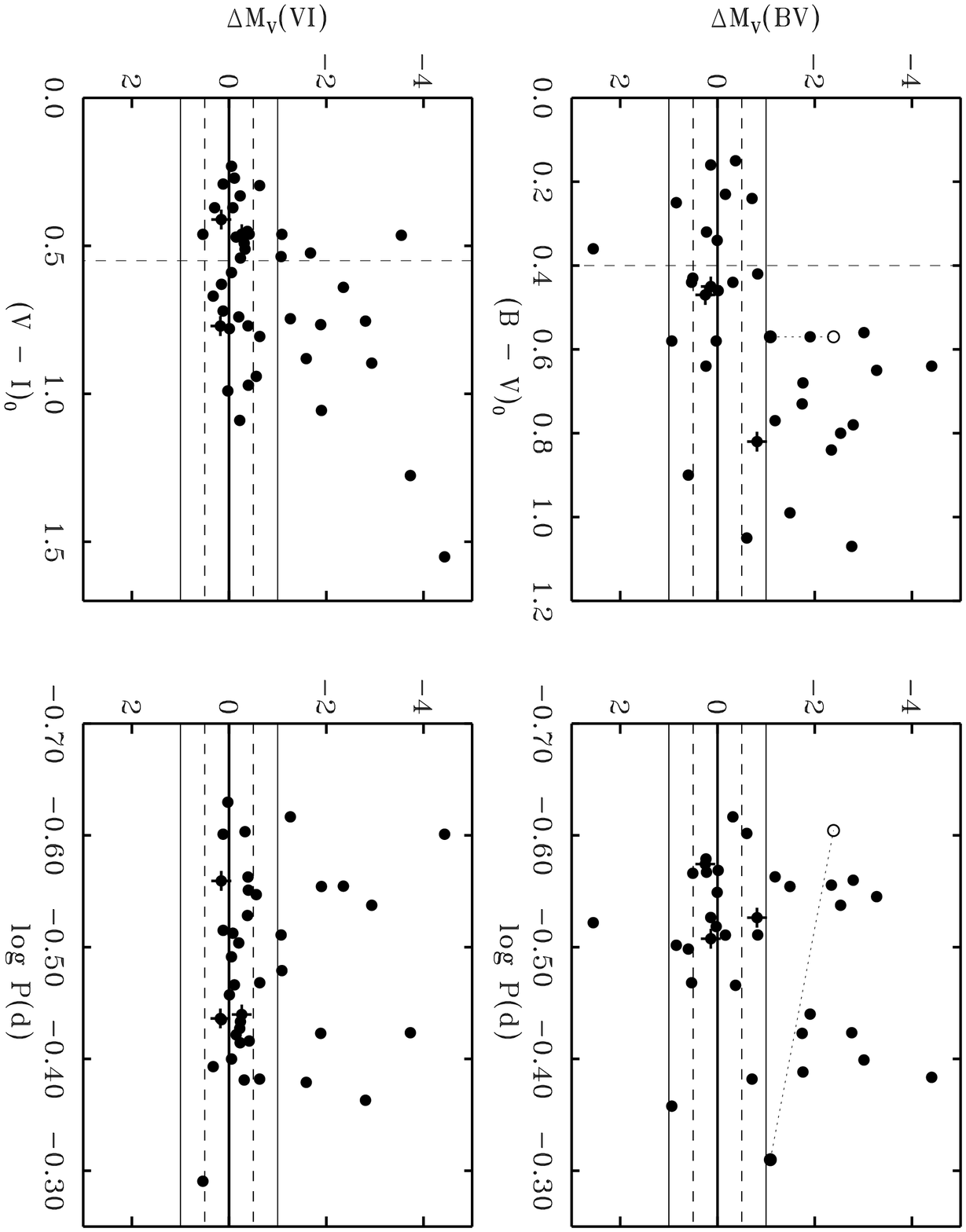]{The absolute-magnitude deviations $\Delta M_V$ are
shown here for both data-sets ($BV$ set in the upper panels, 
$VI$ set in the lower panels), versus the respective colour 
indices and $\log P$ (in days).
Note the large fraction of foreground systems with negative
values of $\Delta M_V$. Most systems
with blue colors are apparently genuine members and also
Blue Stragglers. The approximate locations of the Turn-Off Point
for the ``metal-rich'' GC's are shown by the thin broken
vertical lines (see Figure~5).
The Class-1 members are those with $\Delta M_V < 0.5$ mag.;
the Class-2 members are defined by larger
deviations, but smaller than one magnitude.
\label{fig1}}

\figcaption[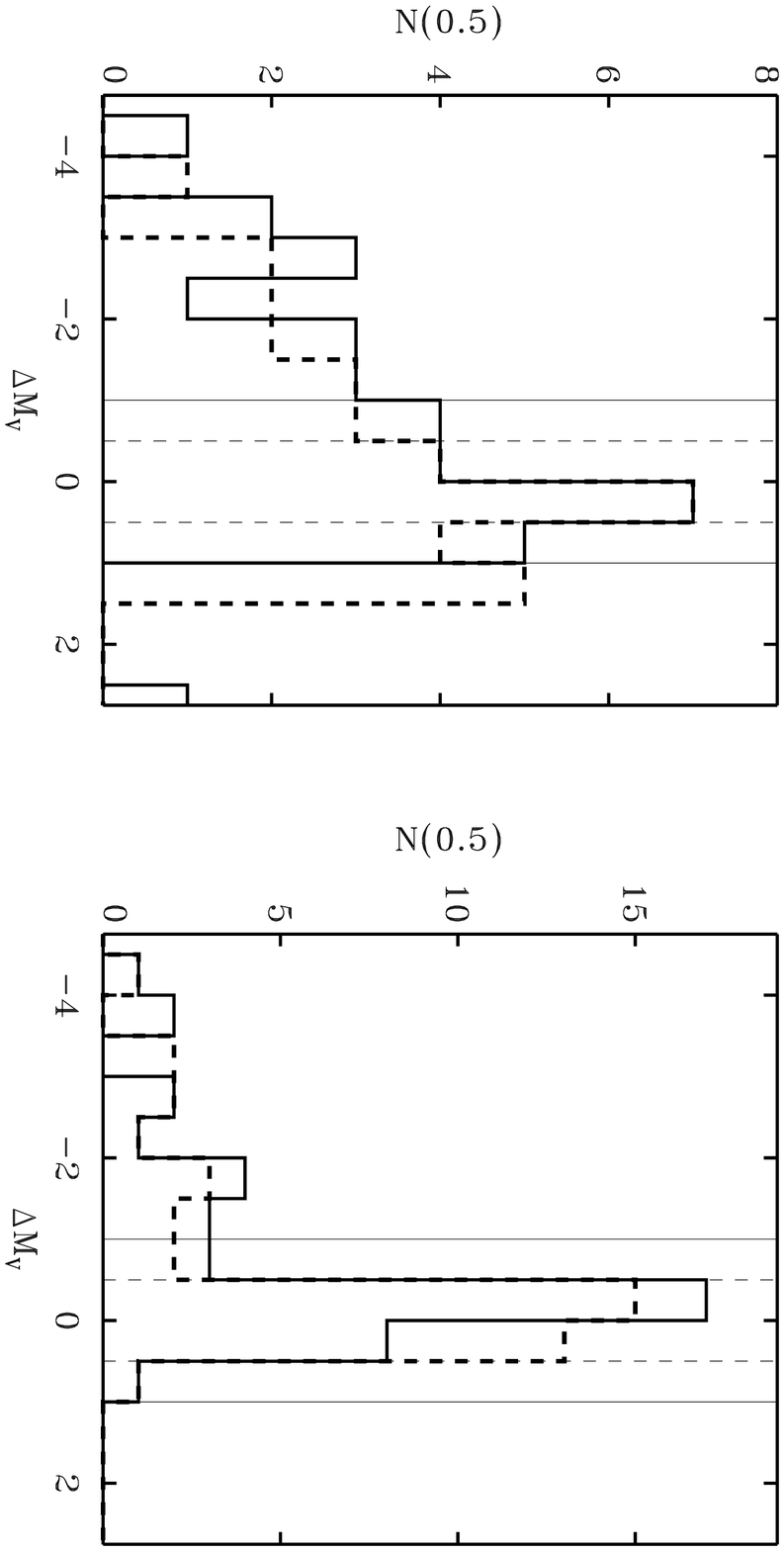]{Histograms of the deviations $\Delta M_V$
with the values of $M_V^{cal}$ computed without 
any metallicity corrections
(continuous lines) and with the metallicity corrections of
$-0.3 \times [Fe/H]$ for the $BV$ set and $-0.12 \times [Fe/H]$
for the $VI$ set added to Equations \ref{eq1} and \ref{eq2}
(broken lines). See the next figure for the
details within a narrow range of the deviations.
\label{fig2}}

\figcaption[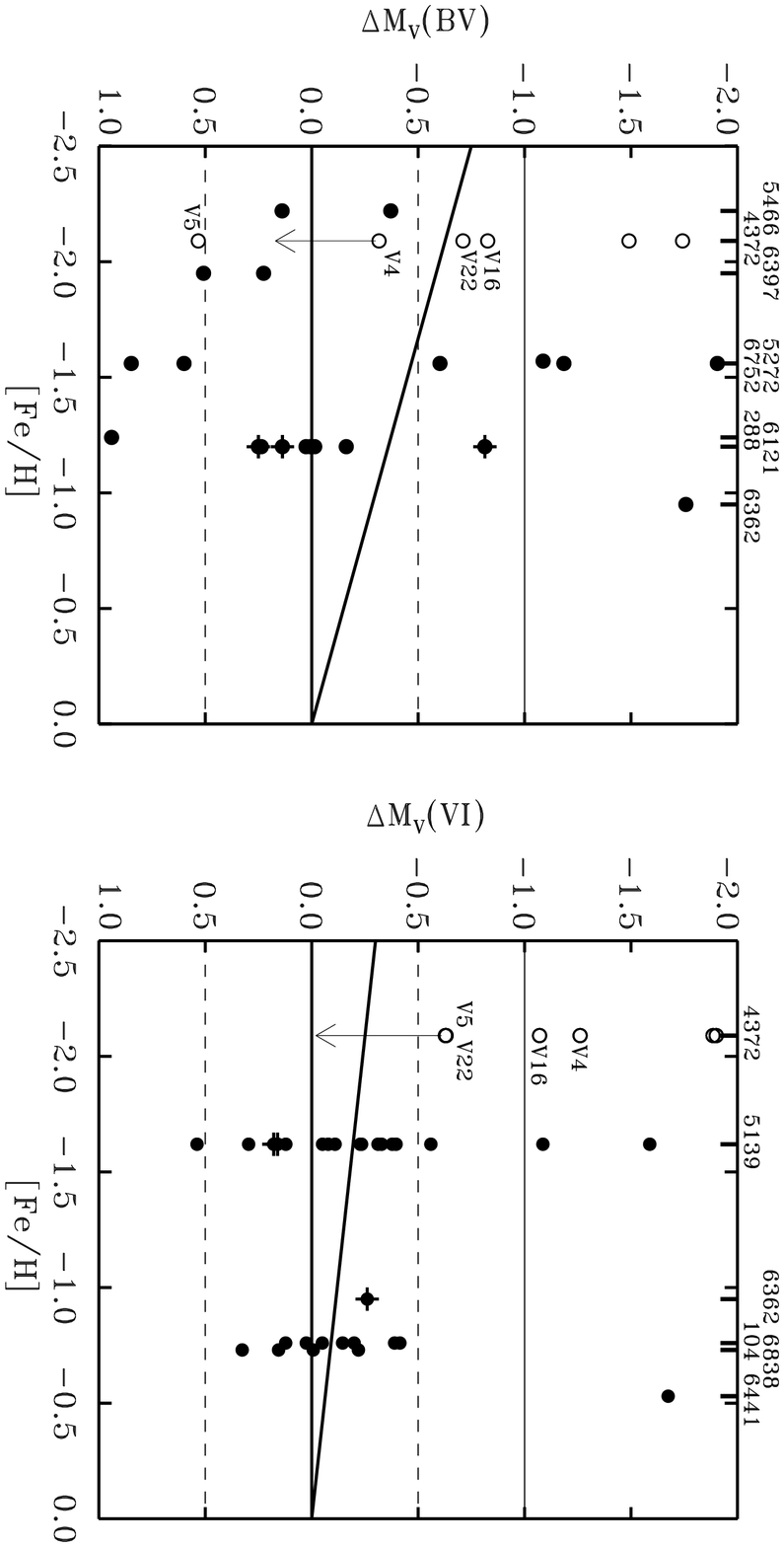]{The band within $-2 < \Delta M_V < +1$ mag.\ 
is shown versus the metallicity parameter $[Fe/H]$. The 
clusters are identified by NGC numbers 
along the upper edges of the panels. The slanting lines 
give the corrections $\propto -0.3\,[Fe/H]$ and $\propto -0.12\,[Fe/H]$. 
The distribution of data
points around the zero-deviation line with and without the
metallicity corrections is the main argument for abandoning
their use in the calibrations. 
The data for NGC~4372 are shown by 
open circles with arrows showing the amount of shift for
all systems in the cluster under 
an alternative set of assumptions on $(m-M)_V$ and $E_{B-V}$
(see the text).
\label{fig3}}

\figcaption[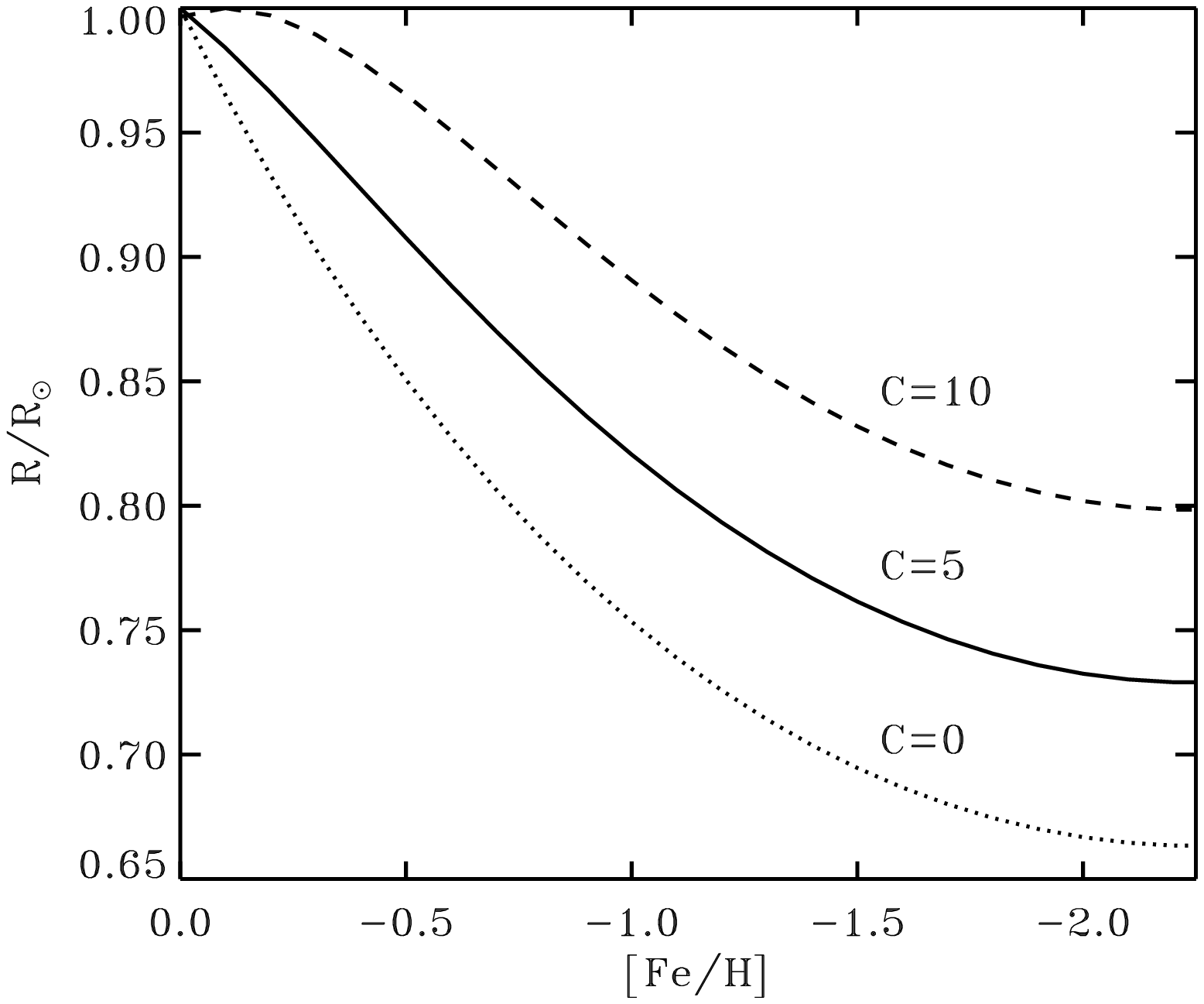]{Radius changes expected from metallicity
variations following the study of Castellani et al.\ 1999.
 The parameter $C = \Delta Y/\Delta Z$
describing the correlation of the helium abundance change with
the heavy-metal abundance change is currently very poorly known.
\label{fig4}}

\figcaption[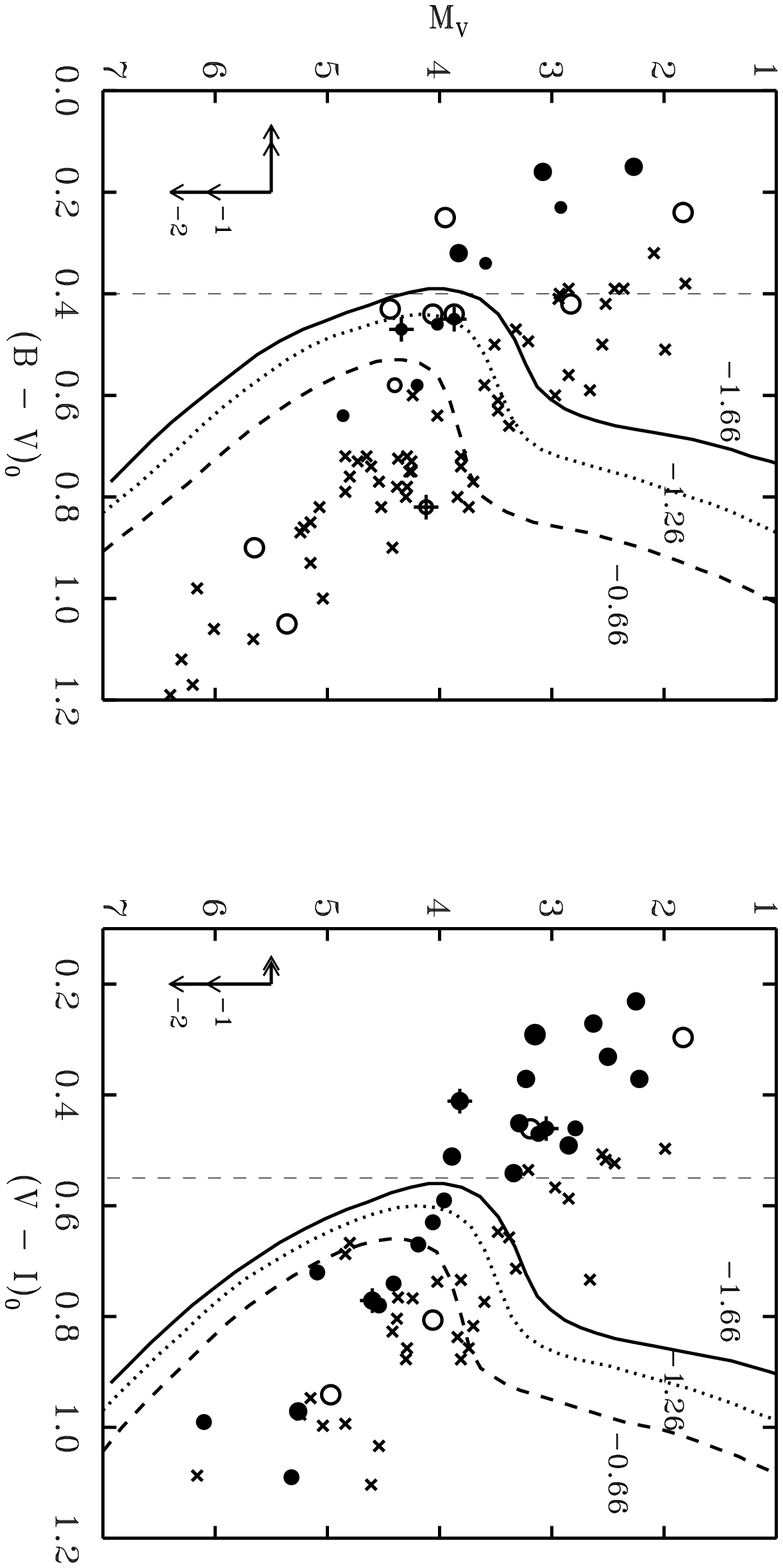]{Color-magnitude diagrams for contact binary
systems in globular clusters. Filled circles mark Class-1 member
candidates which are most probably the real members of the clusters.
Open circles mark Class-2 member candidates which are 
probably not members. Metallicities of individual systems are 
coded by the size of the symbol with
large symbols for metal-poor systems with
$[Fe/H] < -1.5$. The expected metallicity effects are shown 
by arrows in the
left corners of each panel for $[Fe/H] = -1$ and $-2$. 
The vertical broken lines give the approximate limits for
occurrence of Blue Straggler systems at $(B-V)_0=0.4$ and $(V-I)_0=0.55$.
Small crosses show the data for the Disk Population contact systems in 
open clusters (R98); the $(B-V)$ data have been
taken directly from this paper whereas the $(V-I)$ data
are from the original papers cited there.    
\label{fig5}}

\figcaption[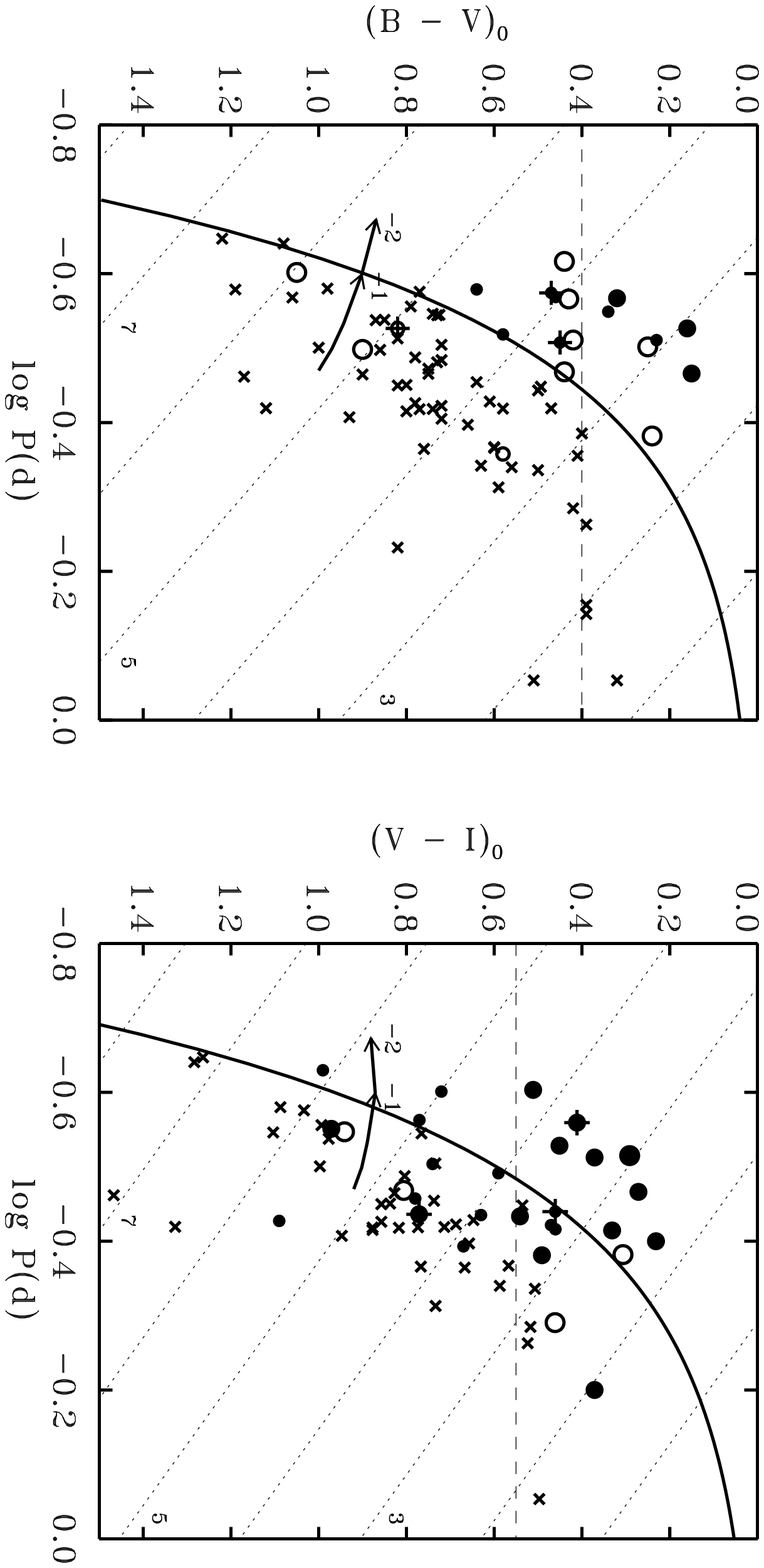]{Period-color diagrams for contact binary
systems in globular clusters. The symbols 
used here are identical as in the
previous figure. The Short Period
Blue Envelopes (SPBE) for the Disk stars are shown as curved lines,
as described in the text. 
The arrows giving the expected variations in the period and color
due to metallicity changes 
for $T_{eff} = 5000$~K have the origin slightly shifted for better
visibility from the nominal values of $(B-V)_0=0.88$ and $(V-I)_0=0.93$.
To give a feel on how absolute magnitudes relate to the plotted
quantities, the slanting dotted lines across the 
figures give the values of $M_V^{cal}$ (not $M_V^{obs}$ as in the
previous figure as these are individual to systems).
\label{fig6}}

\figcaption[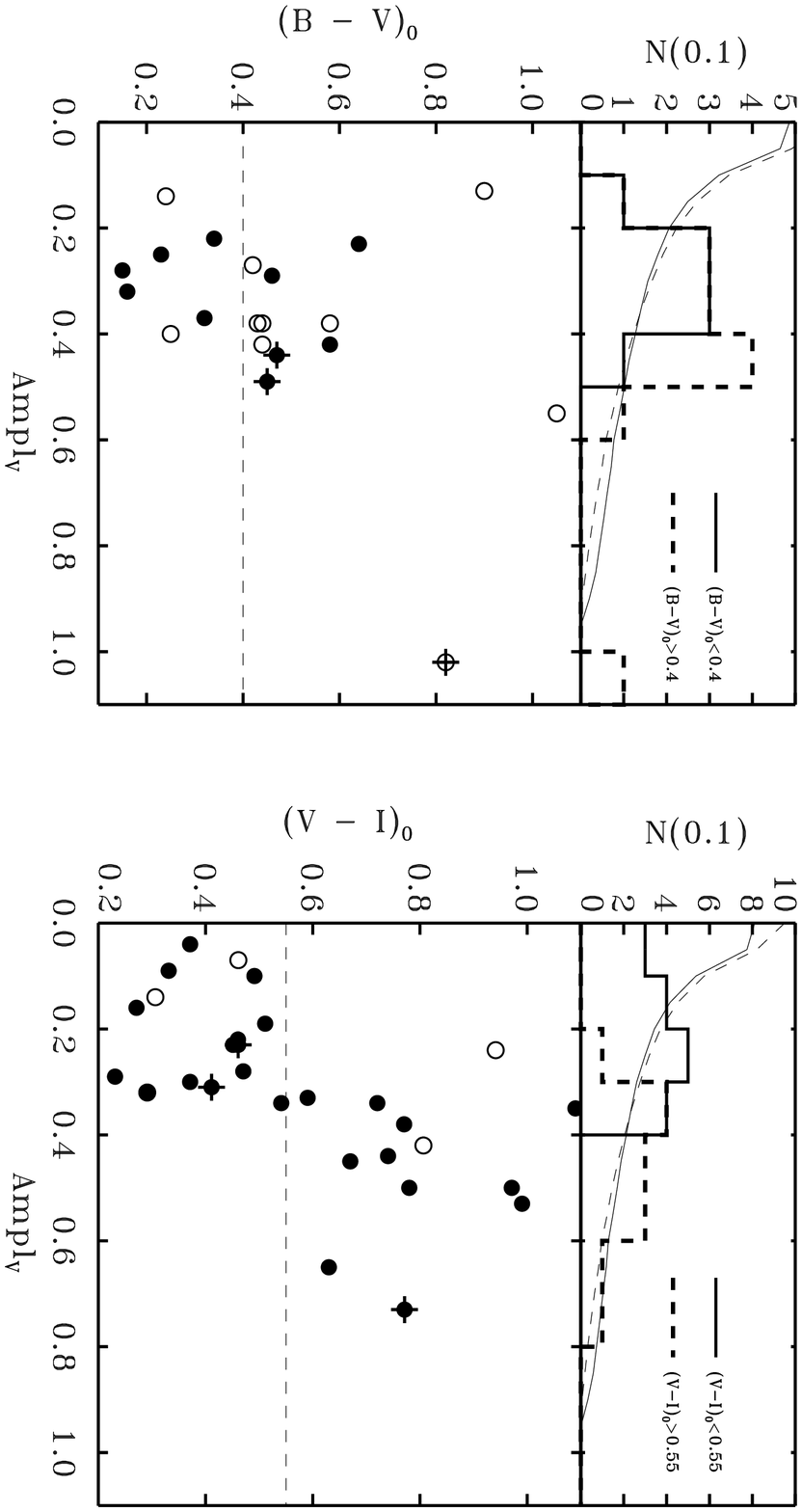]{Amplitudes of light variations for both color-index 
data-sets are shown in relation to de-reddened color indices. 
The vertical broken lines delineate the regions of Blue Stragglers. 
The data with crosses superimposed
on the circles (closed for Class-1 and open for 
Class-2 members) mark the EB-type systems. Since their primary 
eclipses are deeper than for normal contact systems,
they have -- on the average -- larger amplitudes than typical
contact system.
\label{fig7}}

\figcaption[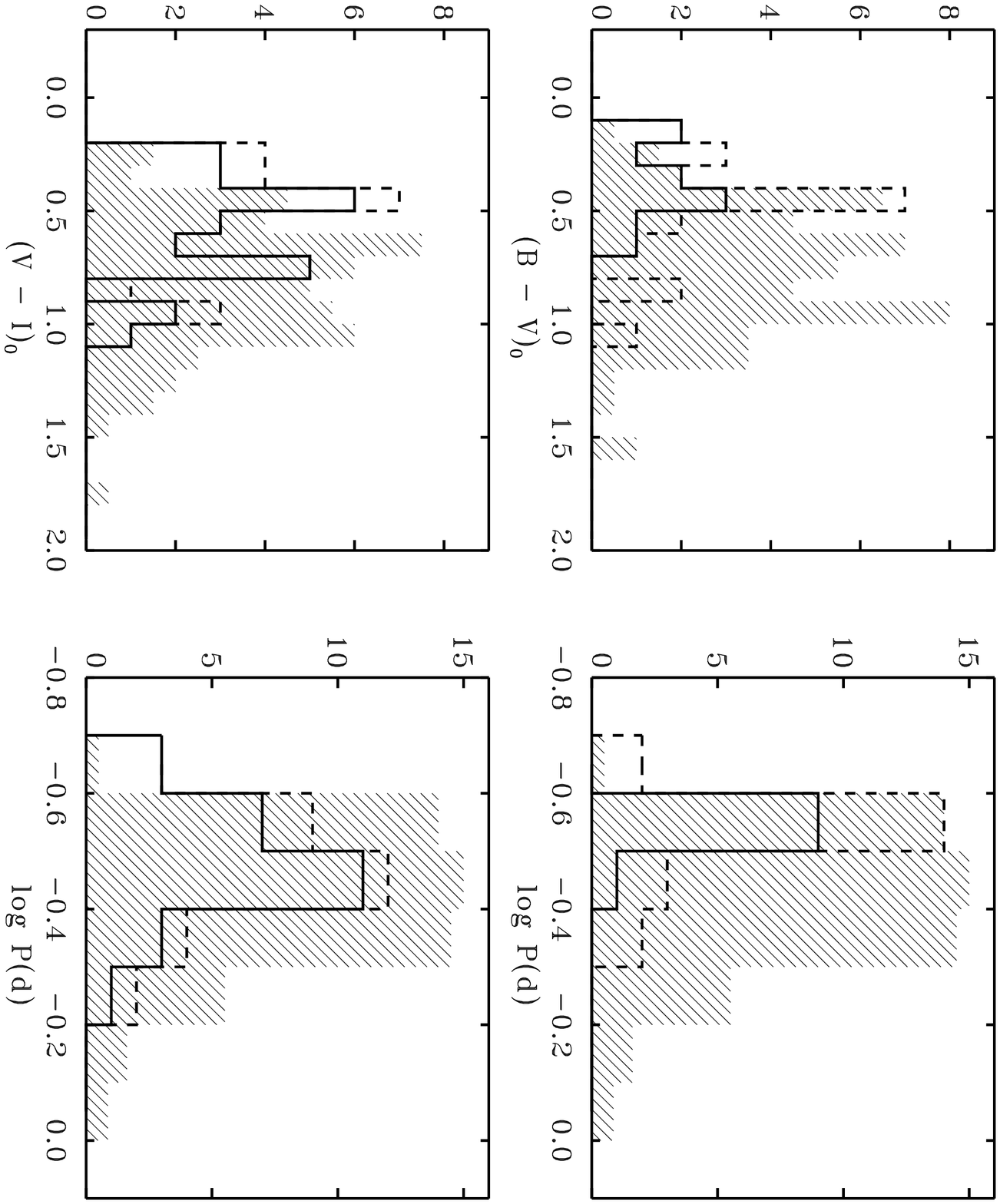]{The color-index distributions for Class-1
members (continuous line histograms) with added Class-2
members (broken line histograms) are compared with Disk
systems from the Baade's Window OGLE sample (R98) (shaded area).
Note that while the OGLE data come from a volume-limited samples
(3 kpc, with modifications at the long-period end from the
5 kpc sample), the GC sample most probably contains 
many detection-selection biases. The OGLE sample data are shown
here with arbitrary normalization, only to indicate the shapes
of the distributions. The upper panels give the $BV$ set data,
the lower panels give the $VI$ set data.
\label{fig8}}

\newpage

\begin{table}                 %
\dummytable \label{tab1}      %
\end{table}

\begin{table}                 %
\dummytable \label{tab2}      %
\end{table}

\begin{table}                 %
\dummytable \label{tab3}      %
\end{table}

\end{document}